\documentclass[useAMS,usenatbib]{mn2e}
\usepackage{graphicx,epsfig,psfig,amsmath,verbatim}
\usepackage{url}
\title[Testing Hydrodynamics Schemes in Galaxy Simulations]{Testing Hydrodynamics Schemes in Galaxy Disc Simulations}
\author[Few et~al.]{C.~G. Few$^{1}$\thanks{E-mail: c.gareth.few@gmail.com},
 C. Dobbs$^{1}$, A. Pettitt$^{2}$ \& L. Konstandin$^{1}$\\
$^{1}$School of Physics, University of Exeter, Stocker Road, Exeter, EX4 4QL\\
$^{2}$Department of Physics, Faculty of Science, Hokkaido University, Sapporo 060-0810, Japan\\
}

\begin{document}
\date{Accepted}
\pagerange{\pageref{firstpage}--\pageref{lastpage}} \pubyear{2016}
\maketitle
\label{firstpage}

\begin{abstract}

We examine how three fundamentally different numerical hydrodynamics
codes follow the evolution of an isothermal galactic disc with an
external spiral potential. We compare an adaptive mesh refinement code
(\textsc{ramses}), a smoothed particle hydrodynamics code (\textsc{sphNG}), and a
volume-discretised meshless code (\textsc{gizmo}). Using standard refinement
criteria, we find that \textsc{ramses} produces a disc that is less vertically
concentrated and does not reach such high densities as the
\textsc{sphNG} or \textsc{gizmo} runs. The gas surface density in the spiral arms
increases at a lower rate for the \textsc{ramses} simulations compared
to the other codes. There is also a greater degree of substructure in the
\textsc{sphNG} and \textsc{gizmo} runs and secondary spiral arms are more
pronounced. By resolving the Jeans' length with a greater number of
grid cells we achieve more similar results to the Lagrangian codes
used in this study. Other alterations to the refinement scheme (adding
extra levels of refinement and refining based on local density
gradients) are less successful in reducing the disparity between
\textsc{ramses} and \textsc{sphNG}/\textsc{gizmo}. Although more similar, \textsc{sphNG} displays
different density distributions and vertical mass profiles to all
modes of \textsc{gizmo} (including the smoothed particle hydrodynamics 
version). This suggests differences also arise which are not intrinsic to
the particular method but rather due to its implementation. The
discrepancies between codes (in particular, the densities reached in the
spiral arms) could potentially result in differences in the locations
and timescales for gravitational collapse, and therefore impact star
formation activity in more complex galaxy disc simulations.

\end{abstract}


\section{Introduction}
\label{sec:intro}

It is well known that galactic dynamics play an important role in the
formation of star-forming regions \citep[e.g.][]{dobbs14}, with the gravitational potential of
the spiral structure competing with hydrodynamical forces, large-scale
differential rotation, and energetic feedback. The
interplay of these processes is particularly difficult to study
observationally and the hydrodynamical complexity means that
a full understanding is analytically intractable. For these reasons,
numerical simulations are a dominant tool for furthering our
understanding of gas dynamics in a galactic context. However
the results can be quite different depending on the particular methodology for
solving these equations. Hydrodynamics codes (which by now are
extremely complex) may give conflicting results due to the respective
strengths and weaknesses of the different implementations. 

Code comparisons seek to quantify how different methodologies 
reproduce fluid flow and where weaknesses lie. These comparisons tend
to concentrate on idealised test conditions \citep{agertz07,tasker08,price08,hopkins15} which are easy to compare
objectively, or on reproducing turbulent behaviour \citep{klessen00,kitsionas09,price10,kritsuk11}, while others
consider cosmological galaxy formation
\citep{frenk99,pearce99,oshea05,torrey12,keres12,scannapieco12}. To
date, almost no work (excepting \cite{hopkins15}) has compared the
behaviour of isolated galaxy discs with different hydrodynamical codes

In this work we compare the Adaptive Mesh Refinement (AMR) code
\textsc{ramses} \citep{teyssier02}, the Smoothed Particle Hydrodynamics (SPH) code
\textsc{sphNG} \citep{benz90}, and the Lagrangian, mesh-less finite volume code
\textsc{gizmo} \citep{hopkins15} in the context of a galactic disc with a
rotating non-axisymmetric potential. Our purpose is to determine whether
spiral galaxy and molecular cloud simulations with these codes are in
concordance, and if not, what measures may be taken to achieve
consistent results. This paper is organised as follows. In the
remainder of \S\ref{sec:intro} we briefly review existing code
comparisons and relevant simulation methods. In \S\ref{sec:method} we
describe the codes employed in this work, our initial conditions and
the set of parameters we use. We analyse the growth of the spiral arms for the
different codes and as a function of resolution in
\S\ref{sec:results}. We conclude with a discussion of our results in \S\ref{sec:discconc}.

\subsection{Hydrodynamical Methods}

\subsubsection{Smoothed Particle Hydrodynamics}

SPH methods use the movement and concentration of particles to provide 
automatic refinement in dense areas. This means that low density regions are 
more poorly resolved and the resolution of phenomena in areas where
density gradients are steep are not guaranteed. Despite these issues,
overdensities are almost always our regions of interest, and SPH
methods are extremely useful.

We will not describe here the details of all SPH codes employed in
this field but simply note works using them \cite[e.g.][]{dobbs06,
  dobbs08c, saitoh08, robertson08, hopkins12, williamson12, grand12, dobbs13, matachavez14,
  williamson14} and that typically resolution is given by particles
numbers of $\sim$10$^6$--10$^7$ and mass resolutions of order
10$^3$--10$^4$~M$_\odot$. Other than deliberate choices of cooling,
star formation, and feedback etc, the key differences between these
codes include how the smoothing length is determined, whether the SPH
equations are used in pressure-energy or density-entropy form, and how
artificial viscosity and conductivity are treated.

\subsubsection{Grid-based Hydrodynamics}

The majority of other work in simulating isolated galactic discs employ some form of grid
method. The simplest of these is a fixed cartesian grid which may fit
around a single spiral arm in the fashion of \cite{kim02, kim06} or
around the entire disc \cite[e.g.][]{wada07,wada08,khoperskov13} for
which spatial resolution depends upon the size of disc that is
simulated but can feasibly reach $\sim$7~pc. Of the works mentioned
here only \cite{khoperskov13} simulates a disc with size comparable to the MW.

To reach greater resolution, grid simulations often use 
adaptive mesh refinement \citep{berger84, berger89}, whereby 
grids are subdivided to some appropriate resolution based on
local hydrodynamical properties. The key strength of this approach is
that it gives the user almost unlimited flexibility and control over
which parts of the simulation volume are to be resolved and that
large scale effects can be incorporated into simulations where very
fine spatial phenomena are to be studied. It should be noted though
that this means that different works can employ different refinement
schemes and parameters which influence the simulations.

One approach requires that the Jeans' length be resolved by a minimum number of cells at each
level. The minimum number is often set to the limit derived in
\cite{truelove97}, i.e. 4 grid cells per Jeans' length. This
refinement criterion is most important when considering
self-gravitating gas and the \cite{truelove97} limit is intended to
prevent numerical fragmentation, although resolving the Jeans' length
is also necessary to capture physical fragmentation of the gas.
The Jeans' length refinement criterion is employed in \cite{tasker09, renaud13, fujimoto14, petit15,
  tasker15}. Of particular note here is the work of \cite{petit15} in which the number of 
grid cells per Jeans' length is increased from the typical 4 to a more rigorous value of 32.

Another approach, sometimes termed `quasi-Lag\-ran\-gian', resolves
cells based on the local density such that the mass enclosed in a grid
cell is roughly the same on each refinement level
\citep[e.g.][]{bournaud10,fujimoto14, agertz15}.
One strength of this technique is that the mass per grid cell may also
include stellar or dark matter mass which are usually gravitationally
dominant. In this way, a stellar substructure can be well resolved for
the gas phase even before the gas density increases. Note that even for
isothermal runs in the absense of non-gaseous mass (such as presented
in this work), a fixed mass threshold for refinement is not equivalent
to a fixed Jeans' length threshold.

The typical finest resolution for AMR runs ranges from 9~pc to 1.5~pc
(considered sufficient to resolve the formation of the largest giant molecular clouds)
with \cite{renaud13} achieving 0.05~pc resolution,
albeit only for the last 50~Myr of the 780~Myr run. However, while oft
quoted as ``the resolution'', stating the minimum grid size is only slightly
more informative than is the minimum smoothing length of an SPH
simulation unless the phenomenon in which one is interested is
entirely comprised of grid cells on that ultimate refinement level. 
Other refinement criteria are available, see \cite{khokhlov98}, but 
those described here cover the approaches currently employed in the simulation of 
galactic discs. 

In addition to the criteria for refinement, and limits
on the permitted levels, refinement usually also takes place in a cubic
buffer around refined cells, this smooths the transition between
different levels and prevents a noisy mesh structure from forming:
this buffer can vary in size and only ever increase resolution,
refining cells that are not ordinarily qualified. Finally, some codes
require de-refinement criteria for the grid, but as \textsc{ramses} is
not among them, we do not discuss this here.

\subsubsection{Hybrid Hydrodynamical Methods}

For a long time the two dominant hydrodynamics schemes were SPH and
grid (often AMR) methods, recently however new techniques have been
developed. One new method, called Godunov-SPH or GSPH, involves a reformulation of SPH to introduce
a Riemann solver that determines the force acting between each particle pair
\citep{inutsuka02,cha03}. This does in principle grant the ability to resolve
shocks in the absense of artificial viscosity and therefore avoid any
side effets that occur as a consequence. 

`Moving-mesh' techniques are a different approach to hydrodynamical
codes which hybridise Lagrangian and Eulerian schemes. This technique
is used in \textsc{arepo} \citep{springel10}. It evolves a
finite-volume unstructured Voronoi mesh that moves with the fluid
flow. This approach retains the ability of Eulerian codes to resolve
shocks by employing a Riemann solver across each boundary between
cells. Movin-mesh schemes are Galilean invariant and less noisy and
less diffusive than SPH codes.
\textsc{arepo} is applied to the problem of an isolated galactic disc
in \cite{pakmor13} and \cite{smith14}.

Another method that hybridises SPH and grid codes is the so-called
`meshless' scheme that is detailed in \cite{lanson08a,lanson08b,gaburov11,hopkins15}.
In these codes, the particle ensemble is topologically similar to moving-mesh frameworks but for
the lack of a sharply defined boundary between the resolution element
domains. These codes are closer in their execution to SPH codes than
moving-mesh techniques are but similarly employ a Riemann solver
across the interfaces between resolution elements, enabling shock
capturing. In this work we consider one of these meshless codes,
\textsc{gizmo} \citep{hopkins15}, in our comparisons.

\subsection{Code Comparisons}

The differences between SPH and grid-based codes have been much
discussed and we now briefly review the most relevant code comparison literature.

The first main type of comparison uses idealised tests that are simple
setups for which analytical solutions exist such as the Sod
shock tube or Sedov blast wave or instability tests
(e.g. Kelvin-Helmholtz). 
\cite{agertz07} compare AMR and SPH using a similar number of
resolution elements in the areas of interest. They found that all grid
codes tested produced similar results for blob
tests, Kelvin-Helmholtz and Rayleigh-Taylor instability
tests and likewise, the SPH codes produced similar results as each
other. \cite{agertz07} find that contact discontinuities and blob
disintegration are less well reproduced by SPH codes, however
\cite{price08} counters that the inclusion of an artificial thermal
conductivity term in SPH formulations reproduces these
discontinuities. Likewise, the introduction of thermal diffusion
improves instability resolution in SPH codes \citep{wadsley08}.
The inclusion of artificial conductivity/thermal diffusion is not relevant to
this particular study because we are considering isothermal flows.

Idealised tests are presented in \cite{tasker08} with the conclusion that SPH codes excel at
resolving the behaviour of dense objects while AMR codes are the
preferred choice for voids (because they retain resolution in low
density regions) and shocks because of the ability to force resolution
in areas of steep density contrast. This work also found that SPH
codes can struggle to resolve fluid instabilities, while praising grid
codes for being able to model multiphase fluids due to the sharp
contrasts that can exist across grid boundaries. Despite these
differences, the conclusion is that concordance between
grid and particle codes is reached when there is one
particle per grid cell in the region of interest.

The second main type of code comparison examines the turbulent
properties of gas under the influence of some driving mechanism. 
\cite{price10} compare the ability of \textsc{phantom} and \textsc{flash} to model supersonic
driven turbulence and achieve similar results with comparable numbers
of resolution elements ($512^3$). Despite similarities,
\textsc{phantom} is better at resolving dense structures, reaching
densities at $128^3$ resolution that \textsc{flash} only achieves at
$512^3$. The resolution of high density regions is best achieved with
SPH, but the grid resolves low density structures better.

In \cite{kitsionas09}, decaying turbulence is modelled with a variety
of static grid and SPH codes. Grid codes are found to be less
dissipative, but for the same number of resolution elements encouragingly similar
results emerge. They conclude that resolution rather than the method
primarily drives differences in turbulence simulations. Supersonic
turbulence decay is also examined in \cite{kritsuk11} as a test of
magnetodynamics codes. While the nature of the simulation set up is not completely
relevant to this work we emphasise the result that even
quite similar codes can generate different results due to minor code
construction choices.

The third common framework for code comparison is galaxy formation
in a cosmological context \citep{frenk99, pearce99,torrey12,keres12, scannapieco12}.
In works of this kind, insights into the accuracy of the hydrodynamical
methods can be masked by dominant gravitational effects or by the method of
analysis which usually focuses on derived physical properties not
directly relevant to the scope of this study. Forthcoming comparisons
in a cosmological context and with an isolated disc is expected under
the AGORA project \citep{kim14}.

Finally, we mention \cite{hopkins15} because in addition to a very
detailed look at a number of idealised tests, the author also provides
a comparative analysis of a cold Keplerian disc test and an isolated
galactic disc, both of which are relevant to the work presented here. 
\cite{hopkins15} shows the results of the viscous instability that
affects SPH realisations due to shear viscosity in the cold Keplerian
disc. Grid methods do not suffer from numerical viscosity in this
context, but advection errors can cause rings to form in what should be
a uniform disc. The work also presents two new methods in the
form of a meshless lagrangrian code either with finite-mass (MFM) or
finite-volume (MFV) resolution elements, which perform well in this test and
others. The analysis of isolated disc galaxy simulations is similar to what we
present in this work. Those runs have lower particle mass
resolution compared to ours but include star formation, stellar feedback, gas
cooling and a live dark matter halo, central black hole, disc and
bulge initialized in equilibrium. The finding here is that SPH runs are similar to 
one another and to the MFM realisations, however MFV transfers mass
outwards due to an angular momentum advection error. It is
particularly noteworthy that the critical flaw seen in SPH when
applied to the Keplerian disc problem (viscous instability) does not manifest in this context
because the pressure forces are much higher and the large stellar/gas mass
ratio allows the stellar component to dominate and stabilise the gas
disc.

The problem with determining the strengths and weaknesses of different
methodologies using idealised tests is that the exact implementation used
in those tests is often changed when the codes are applied to real
problems. One example that stands out is that in idealised tests of
AMR codes, different refinement criteria are applied to that which are
used in production runs. In some cases there is a justification for not applying the same
refinement criteria, i.e. if the phenomenon being studied requires
resolution based on the local density or if the ideal criteria would resolve an impractical amount of the volume
which makes running the simulation intractable. This is the case in
simulations of isolated galactic disc using AMR where
resolution criteria are usually `quasi-Lagrangian' or designed to
resolve the Jeans' length rather than resolving local gradients, as
used for example when performing shock tube or blast wave tests.

\section{Method}
\label{sec:method}

In this work we compare three simulation codes: \textsc{ramses}, \textsc{sphNG}, and 
\textsc{gizmo} in a common framework. We simulate the evolution of isothermal 
gas from an initially uniform surface density disc under the influence of a disc galaxy 
gravitational potential with a fixed spiral perturbation. We now describe each of the 
codes employed in this work.

\subsection{RAMSES}

\textsc{ramses} \citep{teyssier02} is an AMR code in which gas
dynamics are computed with a second-order unsplit Godunov scheme.\footnote{
  \textsc{ramses} is a publicly available code and can be found at\\ \url{http://www.ics.uzh.ch/~teyssier/ramses/RAMSES.html}} This
scheme is inherently shock capturing with no need to invoke artificial viscosity.
Time steps are advanced using a midpoint method with time centered fluxes at cell
boundaries used to update the hydrodynamical variables. The time centered fluxes 
are determined using a second-order Godunov method (otherwise known as the Piecewise 
Linear Method). The duration of the time steps themselves is limited by a modified 
Courant-Friedrichs-Lewy (CFL) condition such that the time step is

\begin{equation}
\Delta t_\mathrm{CFL} = \frac{\Delta x_\ell}{\sum_{i=1}^{N_\mathrm{dim}} (|u_i|+c_\mathrm{s})}\frac{\sqrt{1+2C_\mathrm{CFL}\epsilon_\mathrm{GSR}}-1}{\epsilon_\mathrm{GSR}}
\label{eq:courantram}
\end{equation}

where $\Delta x_\ell$ is the linear extent of a grid cell at level $\ell$ with $N_\mathrm{dim}$(=3) dimensional 
velocity $\boldsymbol{u}$ and sound speed $c_\mathrm{s}$. The right-hand part of equation~\ref{eq:courantram}
replaces the more traditional multiplication by the CFL factor, $C_\mathrm{CFL}$. This is 
changed here to vary with the gravitational strength ratio,

\begin{equation}
\epsilon_\mathrm{GSR} = \Delta x_\ell \frac{\sum_{i=1}^{N_\mathrm{dim}} |g_i|}{\left(\sum_{i=1}^{N_\mathrm{dim}} (|u_i|+c_\mathrm{s})\right)^2}
\end{equation}

where $\boldsymbol{g}$ is the gravitational acceleration experienced by each grid cell.
The right-hand part of equation~\ref{eq:courantram} is equal to $C_\mathrm{CFL}$ 
for $\epsilon_\mathrm{GSR} \to 0$ and is smaller (thus shortening the timestep) 
when the gravitational acceleration is large relative to the local gas velocity.
The \textsc{ramses} runs presented here use $C_\mathrm{CFL}$=0.8. \textsc{ramses} supports different 
time steps for each level of the grid but we have enforced complete synchronicity 
so that the time step of coarse levels is identical to, and thus limited by, lower levels.

\textsc{ramses} is equipped with a number of solver options, the
choices of which are rarely mentioned in literature. In
this work we follow \cite{renaud13} and use the acoustic Riemann
solver with the MinMod slope limiter, however we also include a single run that uses 
the exact Riemann solver and the MonCen slope limiter. We set the mesh-smoothing parameter
$n_\mathrm{expand}$=1.

In this work we use different combinations of refinement criteria
which are now described. The first criterion ensures that the Jeans'
length ($\lambda_\mathrm{J}$) is resolved by $N_{\mathrm{J}}$ cells.
Each cell on level $\ell$ is marked for refinement if

\begin{equation}
\frac{\lambda_{\mathrm{J}}}{\Delta x_\ell} < N_{\mathrm{J}}.
\end{equation}

An alternative criterion forces refinement where the local gradient in a variable $q$ exceeds a
fraction of the value of that variable. Refinement occurs if the
following condition is satisfied,

\begin{equation}
C_\mathrm{q}< 2\cdot\mathrm{MAX}\left[\left|\frac{q_\mathrm{i-1} -
      q_\mathrm{i}}{q_\mathrm{i-1} +
      q_\mathrm{i}+f_\mathrm{q}}\right|,\left|\frac{q_\mathrm{i} -
      q_\mathrm{i+1}}{q_\mathrm{i} +
      q_\mathrm{i+1}+f_\mathrm{q}}\right|\right]
\label{gradeq}
\end{equation}

where $i$-1 and $i$+1 are the cells that bound cell $i$ in each
dimension. The two user defined parameters here are the threshold
$C_\mathrm{q}$ and a pseudo-floor value $f_\mathrm{q}$. In this work
we use only the gradient in density for grid refinement. In principle the
gradient in any hydrodynamical variable may be used, but as our
simulations are isothermal, density and pressure criterion are
degenerate and we have found that for our simulation set up, using 
velocity gradients as a criterion for refinement adds very little if density
gradients are already being used.

One refinement scheme that we do not examine is the
quasi-Lagrangian scheme whereby the mass enclosed in a given cell is
compared to a threshold to determine if the cell is massive enough to
warrant refinement. We do not examine this method because it is
commonly used by combining the mass from the gas phase and stellar
particles, the latter of which are not included in our simulations.

\subsection{\textsc{sphNG}}
\label{sec:sphng}
The SPH code used here (referred to as \textsc{sphNG}) is a modified version of
the code presented in \cite{benz90}. The density of a particle $i$ is estimated 
through a weighted sum of the mass (m) of itself and its neighbours;

\begin{equation}
  \rho_i = \sum_{j} m_j W(|\boldsymbol{x}_i - \boldsymbol{x}_j|,h_{ij})
\end{equation}

where the weighting function where $W$ is the cubic spline kernel which is a 
function of the particle positions $\boldsymbol{x}$ and  
the mean smoothing length for each particle pair, $h_{ij}$. Particles are assigned variable smoothing lengths
according to the local particle density \citep{price04}. The smoothing length and
density are solved iteratively via the Newton-Raphson method according to

\begin{equation}
  h=\eta(m/\rho)^{1/N_\mathrm{dim}}
\end{equation}

where $\eta$=1.2 is equivalent to a typical number of $\sim$58 neighbours for each particle. Artificial viscosity 
is used to capture shocks following \cite{monaghan85} with parameters
$\alpha_\mathrm{v}$=1 and $\beta_\mathrm{v}$=2 \citep[after][]{monaghan92}.

The code employs a second-order Runga-Kutta-Fehlberg integrator 
\citep{fehlberg69} to evolve the hydrodynamics equations.
Particles are assigned individual timesteps, where the timestep of an individual particle 
is the minimum value from several limiters: a conventional CFL condition ($C_\mathrm{CFL}$=0.3) 

\begin{equation}
\Delta t_\mathrm{CFL} = \frac{C_\mathrm{CFL} h}{c_\mathrm{s} + h|\nabla.\boldsymbol{v}| + 1.2(\alpha_\mathrm{v} c_\mathrm{s} + \mathrm{MIN}[0,\beta_\mathrm{v}h|\nabla.\boldsymbol{v}|)]}
\end{equation}

where $\boldsymbol{v}$ is the particle velocity, a force condition which limits the time-step depending on the net acceleration $\boldsymbol{a}$ on a particle, 

\begin{equation}
\Delta t_\mathrm{a} = C_\mathrm{CFL}\sqrt{\frac{h}{|\boldsymbol{a}|}}
\end{equation}

and a third requirement that changes in a particle's velocity, acceleration and smoothing length 
do not exceed a given tolerance, full details are given in \cite{bate95}.

\subsection{GIZMO}

\textsc{gizmo} \citep{hopkins15} is a hydrodynamics code, based on the SPH code
\textsc{gadget-3}, designed to accommodate the benefits of both grid and particle
based hydrodynamics schemes.\footnote{\textsc{gizmo} is a publicly available code and can be found at\\
  \url{http://www.tapir.caltech.edu/~phopkins/Site/GIZMO.html}} The method is based on the works of
\cite{lanson08a, lanson08b} and \cite{gaburov11}, and has
some common features with moving-mesh codes such as \textsc{arepo} \citep{springel10}. \textsc{gizmo} uses a Lagrangian-like formulation, where the
volume is discretised using a weighting function. The weighting function is similar to
the kernel in SPH though in contrast, the kernel gradients play no role
in the equations of motion as they do in SPH. The discretisation of
the fluid is defined by an ensemble of particles that trace the motion
of the cells. Shocks are captured with a Riemann solver, eliminating
the need for artificial viscosity (similar to the approach of Godunov SPH codes). There are
two new methods available in \textsc{gizmo} (as well as two different versions
of SPH with a number of viscosity switches); meshless finite-volume
(MFV) and meshless finite-mass (MFM) methods. The methods differ in
whether the particles/cells are allowed to experience a mass flux
between their neighbours (MFV) or whether their masses are fixed
(MFM). The methods appear to differ only slightly in the test problems
shown in \cite{hopkins15}, especially relative to the differences
seen when compared to pure grid or SPH methods.

The time integration scheme used in \textsc{gizmo} is a second order 
leapfrog integrator very similar to \textsc{arepo} and \textsc{gadget2} 
\citep{springel10} and is described in detail in the appendix G of \cite{hopkins15}. 
Local time-steps are used in all modes so that for each particle the timestep is

\begin{equation}
\Delta t_\mathrm{CFL}  = 2 C_\mathrm{CFL} \frac{h}{|v_{sig}|}
\end{equation}

where $h$ is the kernel smoothing length and $|v_{sig}|$ is the signal velocity 
\citep{whitehurst95,monaghan97,hopkins15}. For the \textsc{gizmo} runs we use a CFL factor of 
$C_\mathrm{CFL}$=0.1, however note that the CFL factor is used differently by each of the 
codes and in particular the way that \textsc{gizmo} and \textsc{sphNG} employ 
this value is not directly comparable with the CFL factor used in RAMSES due to the different 
way in which resolution is defined for each framework. Time-steps in \textsc{gizmo} are also 
limited to prevent spurious events 
of particle interpenetration when neighbouring particles have very different in time-steps \citep{saitoh09,durier12,hopkins14}. 
The time-step may be further restricted depending on the acceleration 
of particles, with $\Delta t_\mathrm{a} = (2 \alpha_\mathrm{k} \epsilon_\mathrm{grav}/|\boldsymbol{a}|)^{1/2}$ after \cite{power03},
where $\epsilon_\mathrm{grav}$ is the force softening length (4~pc) and $\alpha_\mathrm{k}$=0.02.

The MFM and MFV \textsc{gizmo} modes use a standard Harten-Lax-van Leer-Contact (HLLC) Riemann solver 
\citep{toro99,miyoshi05} as the default method. In the rare 
cases where the HLLC solver returns a non-physical result, the code automatically falls back on 
the slower but more accurate exact solver described in \cite{toro97}. Flux-limiting is 
used for the purposes of maintaining numerical stability but we direct the interested 
reader to appendix B of \cite{hopkins15} for a complete and detailed description of how this is 
implemented in \textsc{gizmo}.

One of the SPH modes of \textsc{gizmo} is the traditional density-weighted
approach (such as that of \textsc{gadget2} \citep{springel05}, upon
which \textsc{gizmo} is partly based) which uses a standard artificial
viscosity scheme with no additional measures to allow fluid
mixing instabilities \citep[e.g][]{ritchie01, price08}. While operating as an SPH code, 
a density estimation is required, the density is determined in the 
same way as for \textsc{sphNG} (described in \S\ref{sec:sphng}) but with some 
differences: i) the typical number of neighbours differs with \textsc{gizmo} 
particles having $\sim$32 rather than $\sim$58 and ii) the smoothing length used 
to scale the kernel is simply $h_i$ rather than the mean $h$ of each particle pair.

Another method available in \textsc{gizmo} is `PSPH' \citep{saitoh13, hopkins13}, in which the 
equations of motion are rearranged to combat fluid mixing instabilities, and also includes artificial
conductivity \citep{price08}. In the `traditional' approach the pressure is calculated using the 
density estimate, for PSPH the pressure is instead determined from the neighbouring particles using kernel 
smoothing in the same way as with density.

\subsection{Initial Conditions and External Gravitational Potential}

We initialise our simulations as a uniform disc with a surface density
of 8~M$_\odot$~pc$^{-2}$ and an outer radius of 10~kpc. Gas is
distributed vertically with a sech$^2$($z/H$) profile, where the vertical
scale-height, $H$=0.18~kpc. For our \textsc{ramses} runs we set a density
floor of 6.8$\times$10$^{-32}$~g~cm$^{-3}$. The gas is isothermal and has a
temperature of 1000~K. We neglect self-gravity in order to investigate
the gas response solely to the external gravitational potential. The gas
is initially set up with circular orbits according to the underlying
gravitational potential. The potential is intended to proxy
a rotating stellar mass distribution and is given the logarithmic form
from Binney and Tremaine (1987),

\begin{equation}
\psi(r,z)_{\mathrm{disc}} = \frac{1}{2} v_0^2 \mathrm{log}[r^2 + R_\mathrm{c}^2 + (z/q_\mathrm{\Phi})^2]
\end{equation}

which yields a flat rotation curve with $v_0$=220~km~s$^{-1}$. The radial
and vertical shape of the potential is set by $R_c$=1~kpc and
$q_\mathrm{\Phi}$=0.7. To this potential we add a spiral perturbation of the form given by \cite{cox02}

\begin{multline}
\psi(r,\phi,z) = -4 \pi G H \rho_0 \; \mathrm{exp}\Big(-\frac{r-r_0}{R_\mathrm{s}}\Big)\\
\times\sum_{n=1}^{3} \frac{C_n}{K_nD_n} cos(n\gamma_{\mathrm{s}}) \Big[sech\Big(\frac{K_nz}{\beta_n}\Big)\Big]^{\beta_n}
\end{multline}

where

\begin{equation}
\gamma_{\mathrm{s}} = N \Big[\theta +\Omega_{\mathrm{p}}t -\frac{r/r_0}{tan(\alpha)}\Big],
\end{equation}
\begin{equation}
K_n = \frac{nN}{r sin(\alpha)},\\
\end{equation}
\begin{equation}
\beta_n = K_nH(1+0.4K_nH),\\
\end{equation}
\begin{equation}
D_n = \frac{1 + K_nH + 0.3(K_nH)^2}{1+0.3K_nH},\\
\end{equation}
\begin{equation*}
C_1 = 8\pi/3,\; C_2 = 1/2,\; C_3 = 8\pi/15.\\
\end{equation*}

The parameters $r_0$=8~kpc, $R_{\mathrm{s}}$=7~kpc and $H$=0.18~kpc set the
scaling of the spiral perturbation in three dimensions, $N$=2 is the
number of spiral arms. The pitch angle is $\alpha$=15$^{\circ}$ and
the pattern speed is $\Omega_{\mathrm{p}}$=2$\times10^{-8}$~rad~yr$^{-1}$. The strength of the
spiral perturbation is $\rho_0=\mathrm{m_H}n_{\mathrm{H}}$ with $n_{\mathrm{H}}$=1~atom~cm$^{-3}$.
The effective stellar mass of this potential is $\sim$10$^{11}$~M$_\odot$.

With these parameters the corotation radius is just beyond the edge of
the disc at around 11~kpc. Thus the gas rotation speed 
within the entire disc exceeds that of the spiral perturbation and gas shocks at the
trailing edge of the perturbation.

\begin{figure}
\includegraphics[width=90mm]{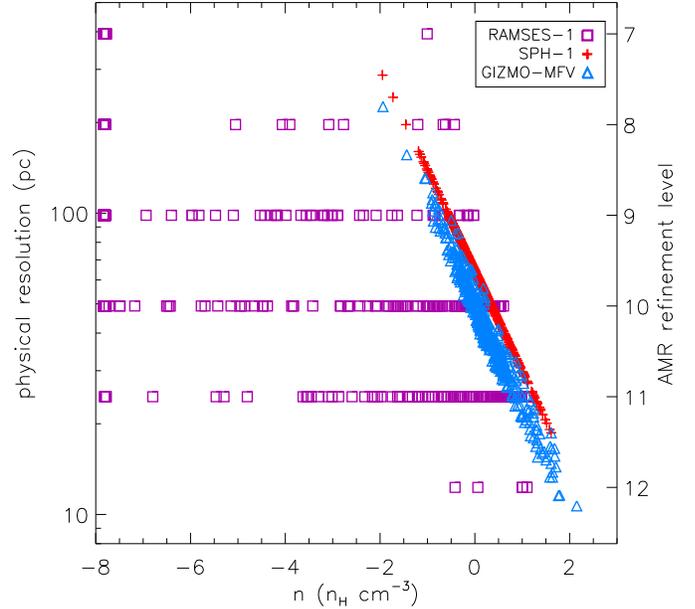}
\caption{Resolution vs. density for the different codes employed in
  this study. Magenta square symbols are \textsc{ramses} leaf cells and the resolution is
  the length of each cell. For \textsc{sphNG} (red cross symbols) and \textsc{gizmo}
  (blue triangular symbols) the resolution is taken as
  twice the smoothing length (see \protect\cite{hopkins15} for an in-depth
  discussion on the use of the kernel length in \textsc{gizmo}).
  For clarity we show only 500 randomly selected 
  resolution elements from each of the simulations.}
\label{fig:rescomp}
\end{figure}

We have made runs using each code but without applying the spiral perturbation to determine if any significant 
features arise within a uniform disc. In this case we note some weak concentric rings; these are not 
related to spurious angular momentum transfer but instead are ripples resulting from the imperfect initial 
pressure equilibrium. We do not think these ripples have any impact on our results for two reasons: 
i) the density contrast of the rings is much weaker than that caused by the spiral potential at all radii and 
ii) we have repeated our baseline runs starting with the disk in vertical pressure equilibrium to reduce the 
impact of the ripples and see no difference in the results.

\subsection{Full Simulation List}

A full list of the simulations used in this work is shown in
Table~\ref{table:parameters} with all the parameters that are
varied. Four of the runs represent our `baseline' models which reflect the choices
made in most of the work on isolated galaxy simulations with external
potentials: these are, \mbox{\emph{RAMSES-1}}, \mbox{\emph{SPH-1}},
\mbox{\emph{GIZMO-MFM}}, and \mbox{\emph{GIZMO-MFV}}. We selected these runs to represent
the typical resolutions found in the literature, 4$\times$10$^6$
particles and 4 grid cells per Jeans' length for particle and AMR runs
respectively.\footnote{\mbox{\emph{RAMSES-1}} does not refine the grid to the
  maximum possible resolution, only reaching level 12 due to the high tolerance of its refinement
  criteria. All other \textsc{ramses} runs make use of the relevant
  maximum level.} We begin with these runs and later move on to discuss
the effect of resolution on a given code.

\begin{table*}
\caption{Overview of the simulation parameters. The top,
  middle and bottom sections describe the \textsc{ramses},
  \textsc{sphNG}, and \textsc{gizmo} runs respectively.
Column (1): simulation reference name; 
column (2): maximum refinement level and corresponding physical size of
smallest grid cell in pc; 
column (3): the number of grid cells or particles at the end of the
simulation (250~Myr);
column (4): number of cells resolving the Jeans' length;
columns (5) and (6): refinement parameters based on density gradients
(see Eq.~\ref{gradeq});
columns (7) and (8): \protect\cite{monaghan85} artificial viscosity parameters;
column (9): Notes for each run, e.g. code type, solvers and viscosity schemes.
}
\begin{flushleft}
\begin{tabular}{l c c c c c c c l}
\hline\hline
name & $\ell_\mathrm{max}$ (pc) & N$_\mathrm{el}$ & N$_\mathrm{J}$ & C$_\mathrm{\rho}$ & f$_\mathrm{\rho}$ & $\alpha_\mathrm{v}$ & $\beta_\mathrm{v}$ & notes\\ [0.5ex] 
\hline
RAMSES-E   & 14 (3.07) & 1.808$\times$10$^6$ & 4   & -    & -     & -  & - & AMR, exact solver + MonCen slope limiter   \\ 
RAMSES-1   & 14 (3.07) & 9.971$\times$10$^5$ & 4   & -    & -     & -  & - & AMR, acoustic solver + MinMod slope limiter   \\ 
RAMSES-2   & 14 (3.07) & 5.362$\times$10$^6$ & 8   & -    & -     & -  & - & AMR, acoustic solver + MinMod slope limiter  \\ 
RAMSES-3   & 14 (3.07) & 2.613$\times$10$^7$ & 16 & -    & -     & -  & - & AMR, acoustic solver + MinMod slope limiter  \\ 
RAMSES-4   & 15 (1.54) & 2.637$\times$10$^7$ & 16 & -    & -     & -  & - & AMR, acoustic solver + MinMod slope limiter \\ 
RAMSES-5   & 12 (12.3) & 3.037$\times$10$^6$ & 4   & 1.3 & 5.3  & -  & - & AMR, acoustic solver + MinMod slope limiter  \\ 
RAMSES-6   & 13 (6.15) & 5.156$\times$10$^6$ & 4   & 1.3 & 5.3  & -  & - & AMR, acoustic solver + MinMod slope limiter  \\ 
RAMSES-7   & 14 (3.07) & 7.676$\times$10$^6$ & 4   & 1.3 & 5.3  & -  & - & AMR, acoustic solver + MinMod slope limiter  \\  
RAMSES-8   & 14 (3.07) & 2.366$\times$10$^6$ & 4   & 1.6 & 5.3  & -  & - & AMR, acoustic solver + MinMod slope limiter   \\
RAMSES-9   & 14 (3.07) & 6.651$\times$10$^6$ & 4   & 1.0 & 5.3  & -  & - & AMR, acoustic solver + MinMod slope limiter   \\ 
\hline
SPH-1         & -             & 4$\times10^6$            & -   & -    & -  & 1      & 2  & SPH + \cite{monaghan85} viscosity\\ 
SPH-2         & -             & 1$\times10^6$            &  -  & -    & -  & 1      & 2  & SPH + \cite{monaghan85} viscosity\\ 
SPH-3         & -             & 8$\times10^6$            &  -  & -    & -  & 1      & 2  & SPH + \cite{monaghan85} viscosity\\ 
SPH-4         & -             & 4$\times10^6$           &  -   & -    & -  & 0.05 & 0.1  & SPH + \cite{monaghan85} viscosity\\
\hline
GIZMO-MFM          & - & 4$\times10^6$            &   -  & -    & -  & -      & - & mesh-less finite mass  \\ 
GIZMO-MFV           & - & 4.593$\times10^6$     &   -  & -    & -  & -      & - & mesh-less finite volume \\ 
GIZMO-MFV-2       & - & 9.188$\times10^6$     &   -  & -    & -  & -      & - & mesh-less finite volume\\ 
GIZMO-MFV-3       & - & 1.131$\times10^6$     &   -  & -    & -  & -      & - & mesh-less finite volume\\ 
GIZMO-SPH-NS     & - & 4$\times10^6$            &   -  & -    & -  & -       & - & SPH + no viscosity switch (constant viscosity) \\ 
GIZMO-SPH-B        & - & 4$\times10^6$            &   -  & -    & -  & -      & - & SPH + \cite{balsara95} viscosity \\ 
GIZMO-SPH-C\&D & - & 4$\times10^6$            &   -  & -    & -  & -      & - & SPH + \cite{cullen10} viscosity \\ 
GIZMO-PSPH          & - & 4$\times10^6$            &   -  & -    & -  & -      & - & PSPH + \cite{cullen10} viscosity\\ 
\hline
\end{tabular}
\end{flushleft}
\label{table:parameters}
\end{table*}

\begin{figure*}
\includegraphics[width=168mm]{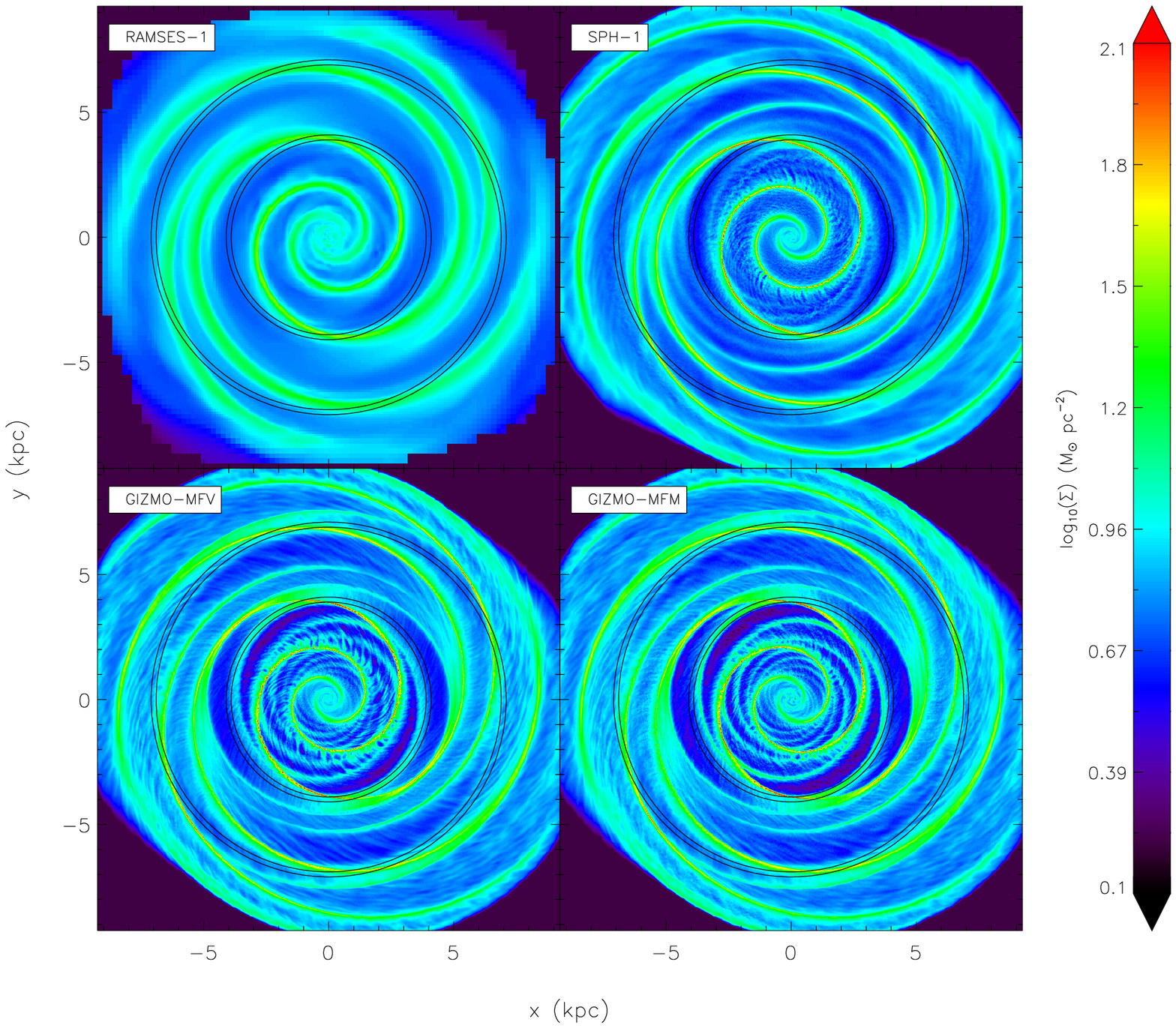}
\caption{Surface density maps for the baseline models at 250~Myr. The panels show \mbox{\emph{RAMSES-1}} 
  (top-left), \mbox{\emph{SPH-1}} (top-right), \mbox{\emph{GIZMO-MFV}} (bottom-left)
  and \mbox{\emph{GIZMO-MFM}} (bottom-right) runs. We overlay black circles
to indicate the location of the annuli at 4$\pm0.1$~kpc and
7$\pm0.1$~kpc that are used for the analysis in section~\ref{armgrowth}. \textsc{sphNG} and \textsc{gizmo} particles are both smoothed over a cubic
spline kernel.}
\label{fig:discimage}
\end{figure*}

Given the difficulty in directly comparing resolutions for grid codes with those
in particle codes, we illustrate the comparative size of the simulation elements
for our baseline simulations in Fig.~\ref{fig:rescomp}, showing the length of grid cells (for a
\textsc{ramses} run) and twice the smoothing length (for the \textsc{gizmo} and \textsc{sphNG}) versus
density.\footnote{Note that the smoothing length in \textsc{gizmo} is
  calculated the same as for an SPH code, but is not used in the
  same fashion for smoothing the particle distribution.}
Our goal here is to compare the simulation codes \emph{as they are
  used in the literature}, but making sure that anything which can overtly
affect the physics of the gas is kept the same, e.g. equation of
state. Fig.~\ref{fig:rescomp} shows that the resolutions of the baseline runs
cover roughly the same range of spatial resolutions except that \textsc{ramses} extends
to much lower densities than the Lagrangian runs. Fig.~\ref{fig:rescomp} also
illustrates that both Lagrangian codes follow a very tight relation between
density and resolution which is not the case with \textsc{ramses}.

\section{Results}
\label{sec:results}

All the simulations presented here start from an initially flat
surface density distribution within the disc region. 
In the absence of self-gravity the ISM responds to the external potential 
and pressure forces leading to an increase in the density of arm
regions over time, however we note that we do not form a steady state
at any point in our simulations. Initially there are only two arms which form around
the spiral perturbation. This is shortly followed by the development
of second pair of arms between the existing ones. These new features
branch from the original arms at $\sim$5~kpc from the center. The appearance
of secondary arm features are well documented as arising from the first ultraharmonic (4:1) resonance 
\citep{shu73, patsis97, chakrabarti03}. We show face-on surface
density maps for the four baseline models in Fig.~\ref{fig:discimage}
in which the original arms and the bifurcating secondary arms are
visible. Both features are most clear for \mbox{\emph{RAMSES-1}} in the upper left panel.

A frequent criticism of grid codes is that angular momentum is not
conserved. For the duration of the runs performed in this work (250~Myr) we
calculate the loss of angular momentum in the AMR runs as $\sim$3\%
and in the SPH and GIZMO runs at $\sim$0.3\%. We do not believe that
angular momentum loss is a significant cause of the differences we find
here, but it may play a more important role in simulations for which
self-gravity results in the formation of small-scale eddies.

\subsection{Overall structure of the disc with different numerical codes}

The different codes tested here respond to the external potential in
slightly different ways. Fig.~\ref{fig:discimage} reveals significant differences between the
\mbox{\emph{RAMSES-1}} and \mbox{\emph{SPH-1}} runs, with less dense arms and far
less fine structure between arms appearing in \mbox{\emph{RAMSES-1}}. 
The \textsc{sphNG} and \textsc{gizmo} runs clearly show the presence of short inter-arm
structures perpendicular to the arms, sometimes referred to as
`spurs' or `feathers', in the inner part of the disc. These features
have been seen in both grid, SPH and \textsc{arepo} \citep{smith14} simulations, and have been attributed
to a number of potential causes including the Kelvin-Helmholz or wiggle
instability \citep{wada04}, feathering instability \citep{lee14}
and orbit crossing in the spiral arms \citep{dobbs06},
although \cite{kim06} suggest purely hydrodynamical instabilities
disappear or are less evident in 3D. Although there is some slight
indication of substructure in the inner part of the \mbox{\emph{RAMSES-1}} disc,
distinct spurs are not visible. The differences between the \textsc{sphNG}
and \textsc{gizmo} runs are largely confined to the strength of the rings
found within the central 3~kpc. The two \textsc{gizmo} runs
are not identical: (\mbox{\emph{GIZMO-MFV}} has less coherent rings than \mbox{\emph{GIZMO-MFM}})
but are very similar to one another when compared with the other two
runs. The growth of rings in a non--self-gravitating
gas disc is also found in grid-based simulations by \cite{shetty06},
in which \emph{leading} spiral structures develop between the arms
near the center of the disc.

\begin{figure}
\includegraphics[width=84mm]{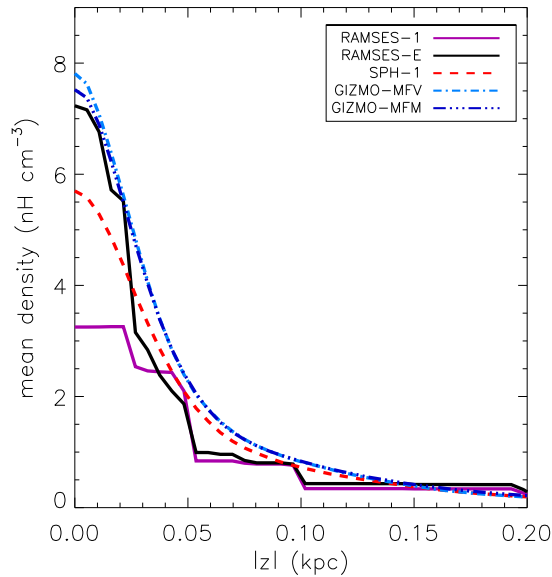}
\caption{Mean density as a function of distance from the disc plane.
  Models \mbox{\emph{RAMSES-1}}, \mbox{\emph{RAMSES-E}}, \mbox{\emph{SPH-1}}, \mbox{\emph{GIZMO-MFV}}, and \mbox{\emph{GIZMO-MFM}}
  are shown as solid magenta, solid black, dashed red, light blue dot-dashed, and dark blue
  triple-dot-dashed lines respectively.}
\label{fig:vertprof}
\end{figure}

Fig.~\ref{fig:vertprof} shows the mean density as a function of
distance from the mid-plane for the four baseline simulations at the
250~Myr mark. In this comparison we also include a run, labelled as \mbox{\emph{RAMSES-E}}, 
which is the same as \mbox{\emph{RAMSES-1}} in all respects except the choice 
of solver and slope limiter. The majority of our \textsc{ramses} simulations 
use quite a diffusive combination of the acoustic Riemann solver with a MinMod 
slope limiter, the \mbox{\emph{RAMSES-E}} run uses the exact Riemann solver with the MonCen slope 
limiter. We use this run to demonstrate the result of using a less diffusive combination.
\mbox{\emph{RAMSES-1}} does not reach the same mass concentration as the Lagrangian runs, having 
only approximately half the mid-plane density of the least concentrated
Lagragian run. The \mbox{\emph{RAMSES-E}} run however has a vertical density profile with a slope 
that is not too dissimilar from the \textsc{gizmo} runs. The two discs realised with \textsc{gizmo} are more 
concentrated than the \textsc{spgNG} run despite having the same particle resolution. 

\begin{figure}
\includegraphics[width=90mm]{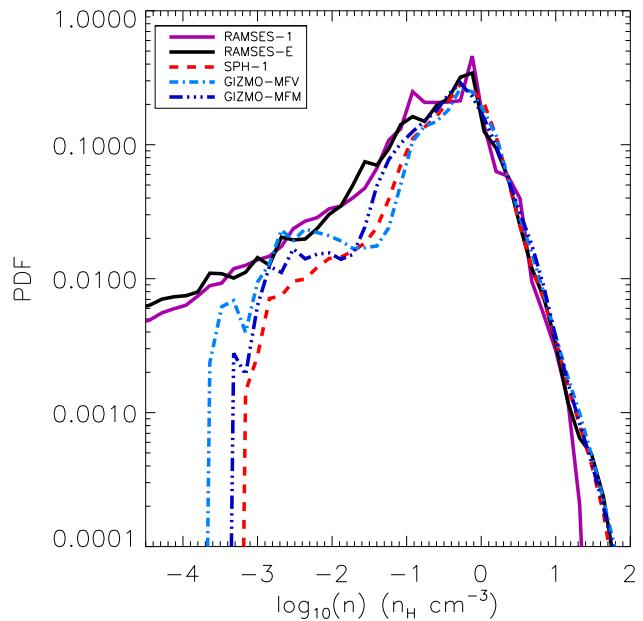}
\caption{Mass-weighted probability density function for the baseline
  simulations. Linestyles are the same as in Fig.~\ref{fig:vertprof}.}
\label{fig:densPDF}
\end{figure}

The one-dimensional structure of the ISM can be examined in the form
of a density probability distribution function (PDF) as plotted in
Fig.~\ref{fig:densPDF}. It indicates a consistent
density peak for all the runs, but differences
do appear in the distributions. The maximum density is curtailed at
10$^{1.3}$~n$_\mathrm{H}$~cm$^{-3}$ in \mbox{\emph{RAMSES-1}} while the
Lagrangian runs extend continuously to around
10$^{1.7}$~n$_\mathrm{H}$~cm$^{-3}$. 
The PDFs of both \textsc{ramses} runs extend down to very low
densities in contrast with the Lagrangian simulations, simply 
because there are resolution elements in the
\textsc{ramses} runs that are not represented by particles in the
other codes. We next describe the evolution of the
spiral arms over time and will discuss the difference between the
codes further in Section~\ref{codediffs}.

\subsection{Growth of Arm Densities}
\label{armgrowth}

\begin{figure}\includegraphics[width=90mm]{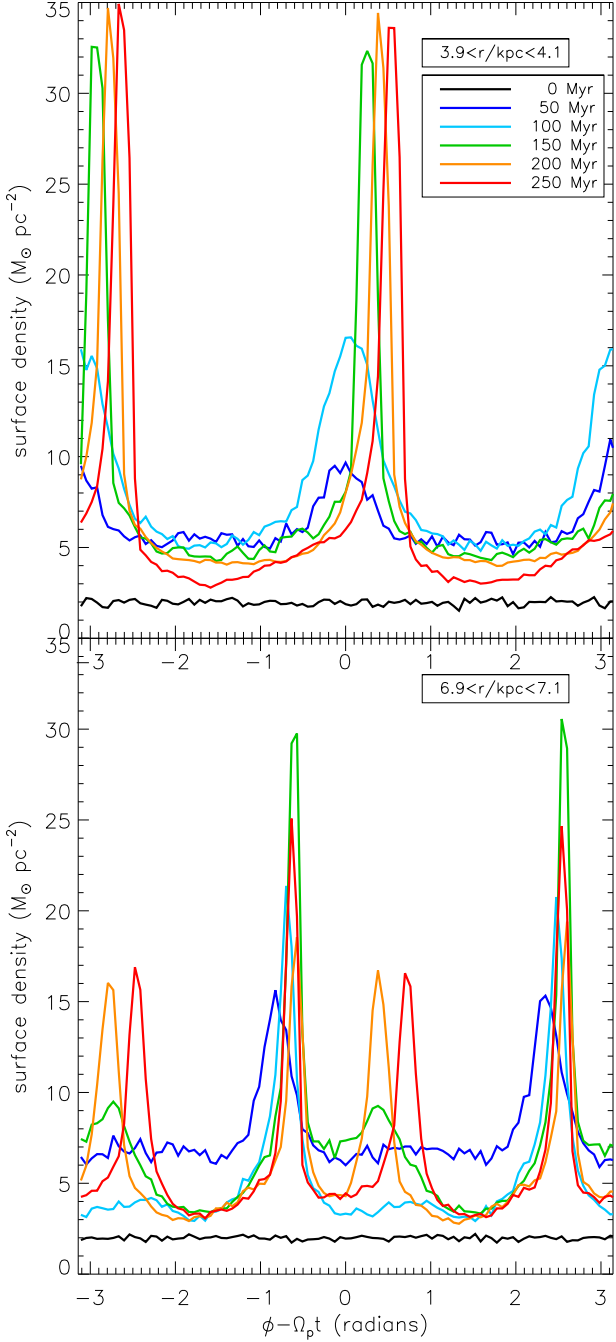}
\caption{Azimuthal profiles of surface density for the \mbox{\emph{SPH-1}}
  run in the rotating frame of reference of the external potential.
  The two panels show different radial annuli (upper: 3.9$<$\emph{r}/kpc$<$4.1
  and lower: 6.9$<$\emph{r}/kpc$<$7.1) and each line represents a different time through
  the run. We choose the two annular rings as one example at a smaller
  radius where no secondary arm emerges and a larger radius where it does.}
\label{fig:timeevo}
\end{figure}

We now consider the surface density of the gas within two annuli, analysing
the azimuthal profile at a number of snapshots throughout the
runs. The time evolution of azimuthal variations in surface density
for the \mbox{\emph{SPH-1}} run is shown in Fig.~\ref{fig:timeevo} for annuli at
4$\pm0.1$~kpc and 7$\pm0.1$~kpc. We show this run as an example which
reflects the typical behaviour of all the runs present here. The
4~kpc annulus (upper panel in Fig.~\ref{fig:timeevo})
illustrates the increasing surface density of the arms until
$\sim$150~Myr, whereupon smaller variations in the peak surface density
are present for the remainder of the simulation. The annulus at 7~kpc
(lower panel in Fig.~\ref{fig:timeevo}) shows the same initial
increase but with larger variations in the peak arm surface density
throughout the run as well as exhibiting the later formation of a secondary
arm feature not seen at smaller radii. The effect of the delay between
the development of the first and second set of arms is to
create oscillations in the maximum surface density. The secondary arms
grow downstream in the gas flow and as its density increases, the density 
of the earlier arm dwindles. This temporarily reduces the maximum surface density until the second arm 
becomes dominant and the maximum surface density increases again, now
representing the secondary arms. 

Oscillations in the density of the arms are also expected due to the
abrupt activation of the spiral potential. In \cite{woodward75} the steepening of spiral waves under the influence
of an arm potential that grows over different timescales is examined, finding that there 
is an initial period of density oscillation which decays over time. \cite{woodward75} also
demonstrate that the more extreme and persistent oscillations occur in the runs where the
spiral potential is activated over shorter timescales. A gradually introduced potential 
reduces the problem of these initial oscillations but for simplicity, and because it is
not common practice to do so, we have not employed such a measure. A gradually introduced
potential does occur naturally in simulations with a live stellar component without initial
spiral structure.

\begin{figure}
\includegraphics[width=84mm]{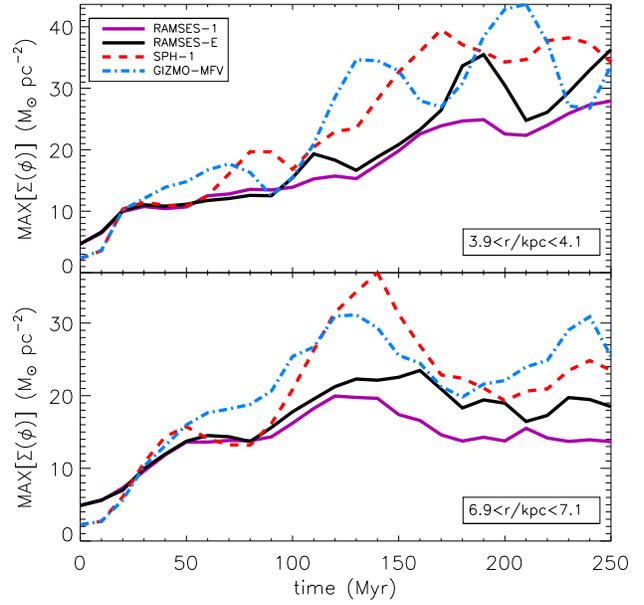}
\caption{Maximum arm surface density for two annuli (4$\pm0.1$~kpc and 7$\pm0.1$~kpc)
  as a function of time. Linestyles are the same as in Fig.~\ref{fig:vertprof}.}
\label{fig:surfdensgrowth}
\end{figure}

Fig.~\ref{fig:surfdensgrowth} shows two annuli at 4$\pm0.1$~kpc and 7$\pm0.1$~kpc
and the maximum surface density within each annulus as a function of
time for each of our baseline models. Fig.~\ref{fig:surfdensgrowth} thus illustrates
the growth of the density of the arms over time. We include only one of the \textsc{gizmo} runs here
(\mbox{\emph{GIZMO-MFV}}) because its counterpart (\mbox{\emph{GIZMO-MFM}}) is extremely similar, see Fig.~\ref{fig:discimage}.
There are substantial differences in the surface density of the arms attained for each of
the codes. In Fig.~\ref{fig:surfdensgrowth} we see that particularly at smaller
radii the \textsc{sphNG} runs differ from the \textsc{gizmo} runs in
the timing and amplitude of the oscillations in arm surface density. 
Lastly we note the considerable difference between \mbox{\emph{RAMSES-1}} 
and the \textsc{sphNG} and \textsc{gizmo} runs, with much lower
densities found in the arm regions and lower amplitude oscillations,
indeed these oscillations are virtually absent in the \mbox{\emph{RAMSES-1}} run. 
When one considers \mbox{\emph{RAMSES-E}} there is an interesting 
difference in the evolution at different radii. At 4~kpc the oscillations are now to some degree apparent and 
the arm surface density is only slightly lower than the runs produced with the other codes, 
however at 7~kpc \mbox{\emph{RAMSES-E}} is not very different to \mbox{\emph{RAMSES-1}} apart from 
a slight increase in the surface density.

\subsection{Dependence of disc evolution on numerical code}
\label{codediffs}

Our comparisons of the baseline galaxy disc simulations highlight a number of
differences between the codes, particularly the \textsc{sphNG} and \textsc{gizmo} runs compared
with \textsc{ramses}. The maximum density in the spiral arms, maximum
density in the midplane and the degree of interarm structure including the
secondary branches are greater with \textsc{gizmo} and \textsc{sphNG},
more so in the \textsc{gizmo} runs.
We now consider which characteristics of the codes might lead to these differences.
The most obvious possibilities are that the codes reach
different effective resolutions, or the inclusion of artificial viscosity in
\textsc{sphNG}. We discuss viscosity next, and describe resolution
tests of the codes in Sections 3.4 and 3.5.

\textsc{sphNG} runs employ artificial viscosity to
allow shock capturing, but artificial viscosity is not required in either
\textsc{ramses} or \textsc{gizmo} as both use a Riemann solver.
For this reason we do not believe that the discrepancy between
\mbox{\emph{RAMSES-1}} and the other runs in Fig.~\ref{fig:surfdensgrowth} 
is due to different viscosity schemes. Additionally we have run
\textsc{sphNG} with reduced viscosity parameters, $\alpha_\mathrm{v}$=0.05 and
$\beta_\mathrm{v}$=0.1. These viscosity parameters yield an arm surface
density growth rate (without oscillations) similar to \mbox{\emph{RAMSES-1}}
but shock capturing is compromised by reducing the artificial
viscosity so harshly.

\textsc{gizmo} can also be operated as an SPH code with various viscosity switches.
In order to further eliminate the influence of viscosity as the cause of differences between
the inviscous MFM and MFV methods of \textsc{gizmo} and the artificially viscous
\textsc{sphNG} we have run our model with a number of these modes (detailed in Table~\ref{table:parameters}).

\begin{figure}
\includegraphics[width=84mm]{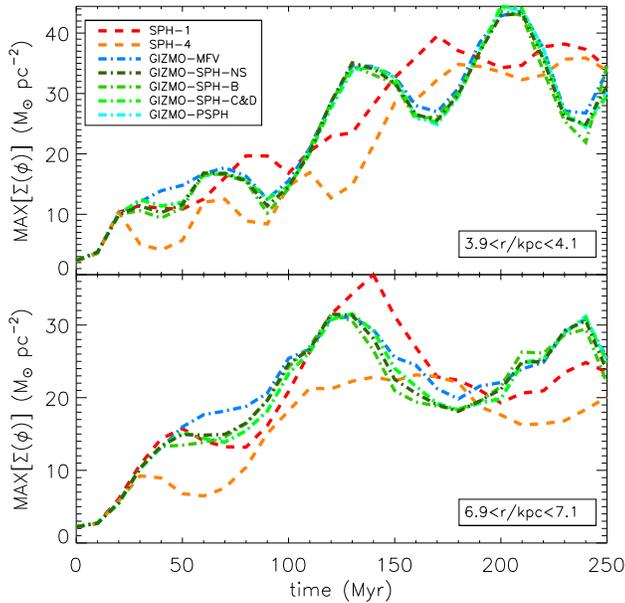}
\caption{Maximum arm surface density for two annuli (4$\pm0.1$~kpc and 7$\pm0.1$~kpc)
  as a function of time for various SPH modes run with \textsc{gizmo} and the baseline 
  \textsc{gizmo} run, \mbox{\emph{GIZMO-MFV}}. We also show \mbox{\emph{SPH-1}} and \mbox{\emph{SPH-4}} to compare 
  the evolution with lower artificial viscosity. Red and orange dashed lines are \textsc{sphNG}, 
  \mbox{\emph{GIZMO-MFV}} is show in blue, and runs and green--cyan lines are runs with \textsc{gizmo} in SPH mode. The 
\mbox{\emph{GIZMO-PSPH}} almost exactly follows the \mbox{\emph{GIZMO-SPH-C\&D}} run.}
\label{fig:surfdensgrowthsphmode}
\end{figure}

These models are compared with \mbox{\emph{GIZMO-MFV}} and \mbox{\emph{SPH-1}} in Fig.~\ref{fig:surfdensgrowthsphmode}.
It is clear that the artificial viscosity scheme makes little difference and all \textsc{gizmo}
modes behave in roughly the same way. Therefore, the cause of the discrepancy found between the \textsc{sphNG} and \textsc{gizmo} runs is 
not a basic difference in the SPH or the meshless Lagrangian methods of \cite{hopkins15}.
The fact that, despite some minor differences, the evolution of the \textsc{gizmo} SPH runs are 
far more similar to \mbox{\emph{GIZMO-MFV}} than they are to \mbox{\emph{SPH-1}} means that it is likely that 
some aspect of the code is responsible for the offset between the two codes that is separate from 
the fundamental methodologies. We have not explored the codes in sufficient detail to offer a definitive explanation of this
but note that, apart from the core hydrodynamics solver, all components of \textsc{gizmo} are commons to both the 
MFV/MFM and SPH modes. The different results found with \textsc{sphNG} and \textsc{gizmo} may therefore be due to differences in the neighbour
finding process, the time-step criteria or time integration.

We note that the SPH kernel for \textsc{gizmo} and \textsc{sphNG} is a cubic spline but that the smoothing
length is defined differently with the smoothing lengths in \textsc{gizmo} being smaller (see
Fig.~\ref{fig:rescomp}). Exploring this, we have rerun \mbox{\emph{GIZMO-MFM}} and \mbox{\emph{GIZMO-SPH-C\&D}} with
the same particle smoothing lengths as our \textsc{sphNG} runs and find that it makes very little difference to the arm surface
density and density PDF, however the vertical concentration is slightly reduced and therefore closer to
the profile of \mbox{\emph{SPH-1}}. We also observe a slight weakening of interarm structures when using
larger smoothing lengths.

\subsection{Resolution of \textsc{sphNG} and GIZMO Runs}

\begin{figure}
\includegraphics[width=90mm]{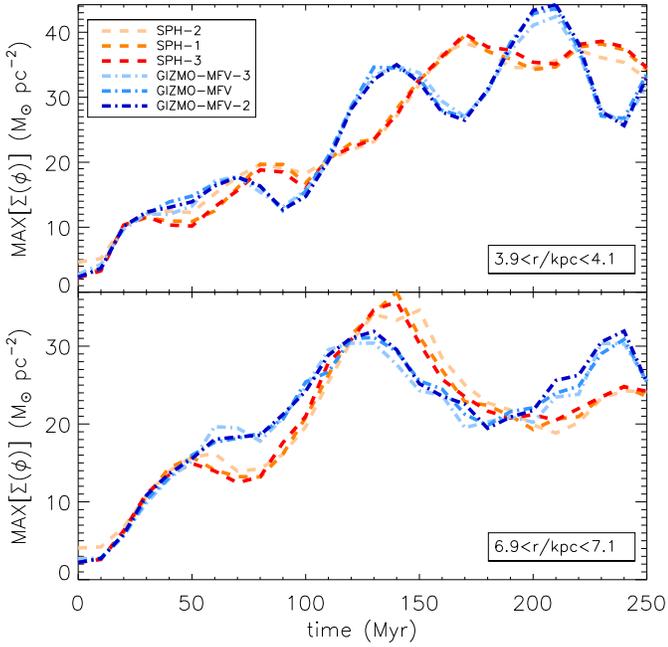}
\caption{Maximum arm surface density as a function of time for
  \textsc{sphNG} (red/orange dashed lines) and \textsc{gizmo} (blue dot-dashed lines) runs
  with different resolutions (initially with 1, 4 and 8 million particles). Darker colours represent
  greater resolutions.}
\label{fig:surfdensgrowthreslagrange}
\end{figure}

In this section we consider how our two Lagrangian codes (\textsc{sphNG} and \textsc{gizmo}) evolve differently with varying mass
resolution. The evolution of the arm surface density for runs that
initially have 1$\times$10$^6$, 4$\times$10$^6$, and 8$\times$10$^6$ particles
are shown in Fig.~\ref{fig:surfdensgrowthreslagrange}. The number of
particles in \textsc{sphNG} and \textsc{gizmo}-MFM runs is fixed, but the
\textsc{gizmo}-MFV runs allow particle splitting and increase the
number of particles over the course of the simulation. The final number of particles
for \mbox{\emph{GIZMO-MFV-3}}, \mbox{\emph{GIZMO-MFV}}, and
\mbox{\emph{GIZMO-MFV-2}} are 1.131$\times10^6$, 4.593$\times10^6$, and
9.188$\times10^6$ respectively.

For both of the Lagrangian codes, the evolution of the peak surface
density is invariant with resolution, i.e. the rate of arm growth is
unaffected. \textsc{sphNG} and \textsc{gizmo} runs do not show much
difference in the surface density of the galaxy arms as a function of resolution, but we
do find differences between the codes themselves. Despite the codes
presenting a similar time-averaged growth curve, the oscillations
discussed in the previous section are offset in time (see Fig.~\ref{fig:surfdensgrowthreslagrange}).

\begin{figure}
\includegraphics[width=90mm]{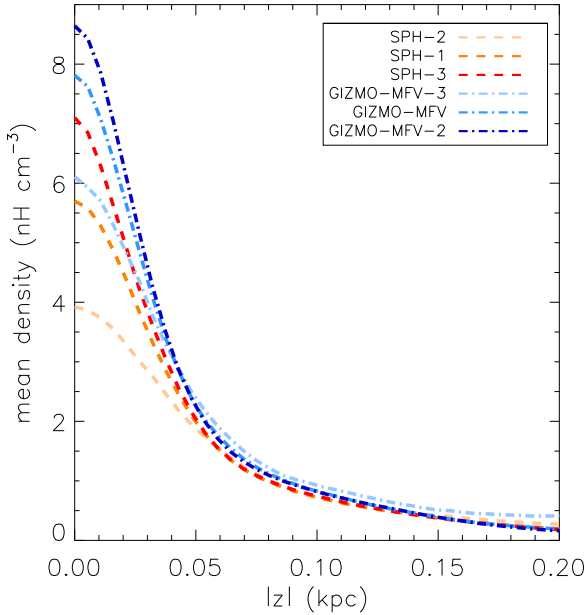}
\caption{The vertical density profiles of \textsc{sphNG} (red/orange dashed lines) and \textsc{gizmo}
  (blue dot-dashed lines) runs with different resolutions (initially
  with 1, 4 and 8 million particles). Darker colours representing greater resolutions.}
\label{fig:lagrangeresvertdens}
\end{figure}

Despite the invariance of the maximum surface density with resolution
we do see that the vertical density profiles are steeper for runs with
higher resolution, as shown in Fig.~\ref{fig:lagrangeresvertdens} and
consequently the peak volume density is enhanced by resolution. 
As the particle resolution is increased (and the smoothing length
shortens) the vertical density profiles is improved.
The \textsc{gizmo} runs exhibit steeper density profiles
than do the \textsc{sphNG}, this is partially due to the different definition and
use of the smoothing length/particle domain.

\begin{figure}
\includegraphics[width=90mm]{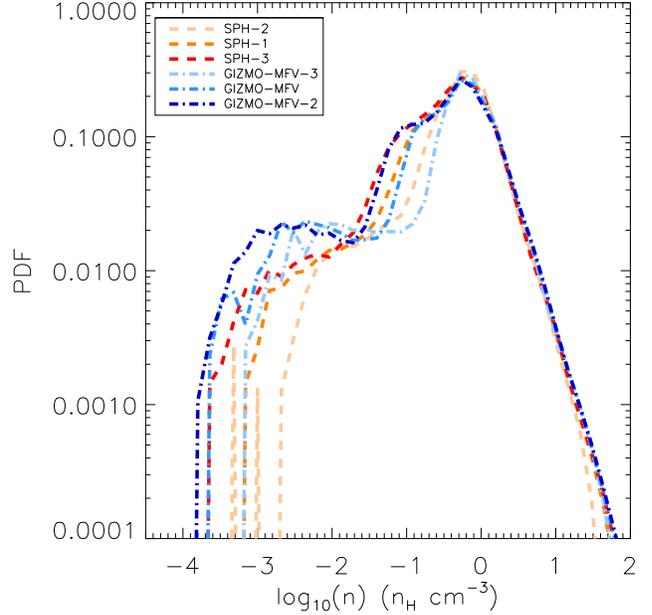}
\caption{Mass-weighted probability density function for Lagrangian simulations
  with different resolutions. Lines show \textsc{sphNG} (red/orange dashed lines)
  and \textsc{gizmo} (blue dot-dashed lines) runs that initially have 1, 4 and 8 million particles,
  darker colours represent greater resolutions.}
\label{fig:denspdflagrange}
\end{figure}

The mass-weighted density PDFs for our resolution comparison of Lagrangian runs
are shown in Fig.~\ref{fig:denspdflagrange}. For both \textsc{sphNG} and
\textsc{gizmo}, increased resolution broadens the distribution and increases the
fraction of gas at lower densities. \textsc{gizmo} runs have a consistently narrower
distribution in the high density peak but extend to lower densities than their \textsc{sphNG}
counterparts. The maximum density achieved by all the runs is quite consistent with
the sole exception of \mbox{\emph{SPH-2}} (the lowest resolution SPH run) which is truncated
around 0.2 dex below the others.

\subsection{Resolution of AMR runs}
\label{AMRressec}

\begin{figure}
\includegraphics[width=90mm]{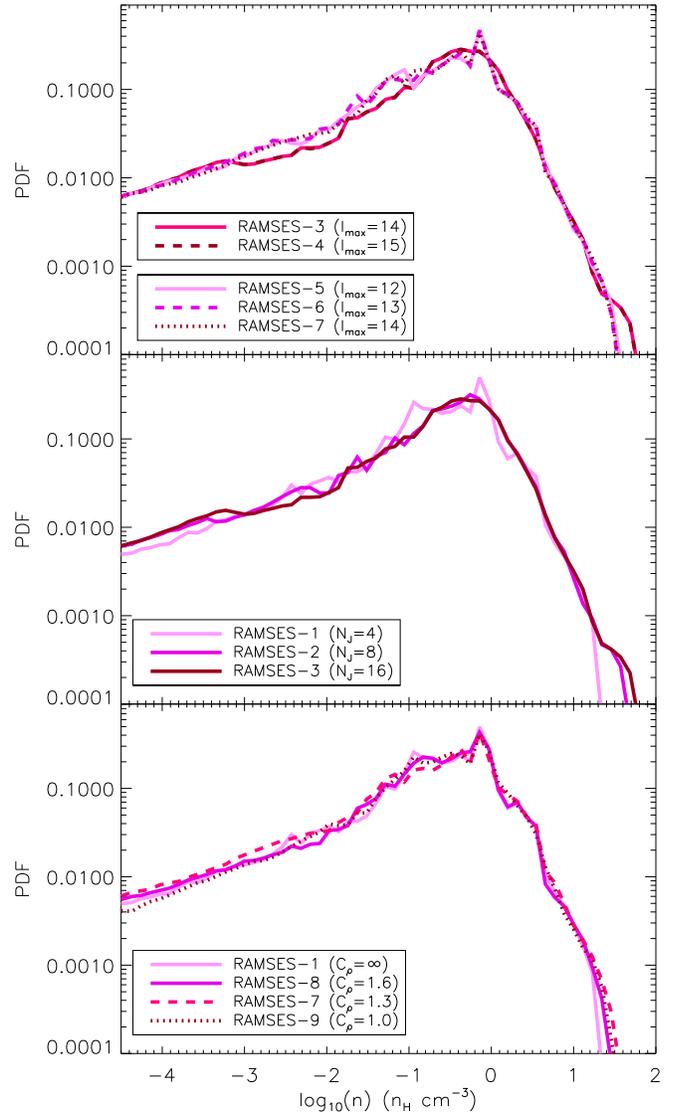}
\caption{Mass-weighted probability density functions for \textsc{ramses} runs
  with different refinement schemes. In the top panel we compare runs
  with differing values of $\ell_\mathrm{max}$. The middle panel compares
  different $N_\mathrm{J}$ values. The lower panel shows runs with
  different density gradient refinement thresholds ($C_\rho$). Darker colours
  represent greater resolutions: see Table~\ref{table:parameters} for details
  of the parameters used in each run.}
\label{fig:denspdfram}
\end{figure}

\begin{figure*}
\includegraphics[width=180mm]{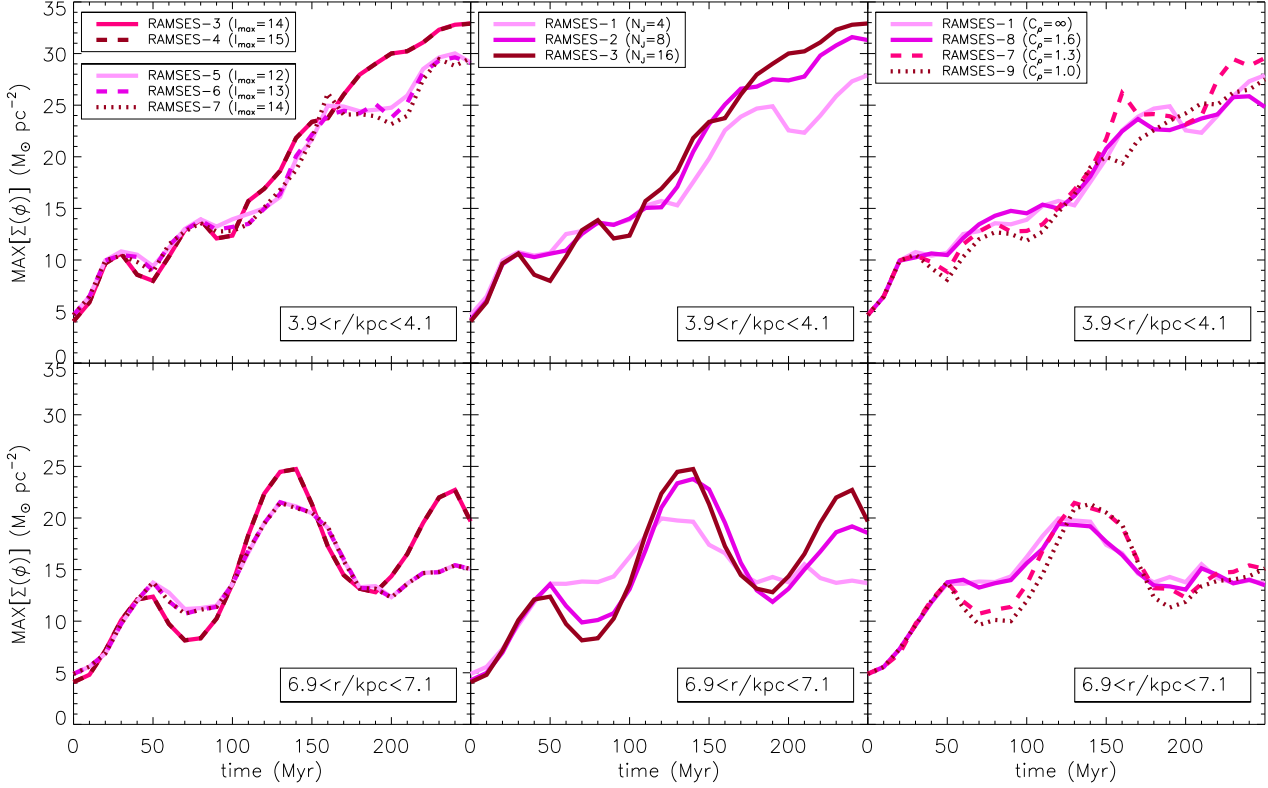}
\caption{Maximum arm surface density as a function of time for
  \textsc{ramses} runs with various refinement schemes. The upper
  panels illustrate the evolution of the spiral arms at 4$\pm0.1$~kpc
  and the lower panels at 7$\pm0.1$~kpc. The left-hand panels compare runs
  with differing values of $\ell_\mathrm{max}$. The middle panels show
  runs with different $N_\mathrm{J}$. The right-hand panels show runs with
  different density gradient refinement thresholds ($C_\rho$). Darker
  colours represent greater resolutions: see Table~\ref{table:parameters}
  for details of the parameters used in each run.}
\label{fig:surfdensgrowthresram}
\end{figure*}

Resolution in AMR simulations is not a linear characteristic. We can change
the minimum and maximum refinement levels or the parameters governing
refinement. In this section we compare a number of approaches
to varying the resolution within \textsc{ramses}. In addition to
the previously shown run (\mbox{\emph{RAMSES-1}}) which employs a Jeans'
length refinement criterion with the typical threshold
$N_\mathrm{J}$=4 we now test the effect of varying $N_\mathrm{J}$ and
$C_\rho$, which control grid refinement according to the local Jeans' length
and density gradients respectively. We also vary $\ell_\mathrm{max}$
which is the upper limit on the grid level. The specific parameters used
in each run are detailed in Table~\ref{table:parameters}. We now
discuss these three parameters that control refinement in turn using
the mass-weighted density PDFs in Fig.~\ref{fig:denspdfram} and the maximum surface density within
annuli at 4$\pm0.1$~kpc and 7$\pm0.1$~kpc as a function of time in
Fig.~\ref{fig:surfdensgrowthresram}. 

Firstly, we consider the effect of increasing $\ell_\mathrm{max}$
which permits the code to refine the grid to higher levels. For this
we direct the reader to the runs shown in the top panel of
Fig.~\ref{fig:denspdfram}, i.e. we compare \mbox{\emph{RAMSES-3}} and
\mbox{\emph{RAMSES-4}} which have $\ell_\mathrm{max}$= 14 and 15 respectively
but with all other refinement criteria the same. We also compare three runs
that have an alternative set of refinement criteria to the previous two 
which use $\ell_\mathrm{max}$= 12, 13 and 14 (\mbox{\emph{RAMSES-5}},
\mbox{\emph{RAMSES-6}} and \mbox{\emph{RAMSES-7}}), see Table~\ref{table:parameters} for details.
We see here that increasing $\ell_\mathrm{max}$ does not have an
enhancing influence on these simulations, indeed \mbox{\emph{RAMSES-3}} and
\mbox{\emph{RAMSES-4}} have identical density PDFs. We further 
examine the growth of the spiral arm surface density in the
left-hand panels of Fig.~\ref{fig:surfdensgrowthresram}. Again we note
that varying $\ell_\mathrm{max}$ makes almost no difference to the
evolution of the galactic disc, although in this case there is a marginal
reduction in arm surface density as $\ell_\mathrm{max}$ increases for
both the 4 and 7~kpc annuli in the case of \mbox{\emph{RAMSES-5}}, \mbox{\emph{RAMSES-6}} and \mbox{\emph{RAMSES-7}}.

We now consider whether increasing the number of grid cells that resolve
the Jeans' length ($N_\mathrm{J}$) has an influence on the
simulations.  \mbox{\emph{RAMSES-1}}, \mbox{\emph{RAMSES-2}}, and \mbox{\emph{RAMSES-3}}
have $N_\mathrm{J}$ = 4, 8 and 16 respectively and are shown in the
middle panel of Fig.~\ref{fig:denspdfram}. We note a marked increase
in the maximum density achieved when $N_\mathrm{J}$ takes greater values.
In fact the densest end of the density distribution function for
\mbox{\emph{RAMSES-3}} (for which $N_\mathrm{J}$=16) is consistent with that
found in the \textsc{gizmo} and \textsc{sphNG} runs. The central panels of
Fig.~\ref{fig:surfdensgrowthresram} illustrate how increasing
$N_\mathrm{J}$ alters the growth of the spiral arms. We see that
greater values of $N_\mathrm{J}$ produce higher surface densities at
4~kpc. At 7~kpc we see simply that as $N_\mathrm{J}$ increases, the
oscillations that are clearly present in \textsc{sphNG} and \textsc{gizmo} runs
(see Fig.~\ref{fig:timeevo}) become more apparent. 

One key strength of grid codes is the ability to resolve sharp density
contrasts, achieved partially through the use of refinement criteria
based on the local gradient in hydrodynamical variables. The final
comparison we make varies the threshold that controls grid refinement
based on the density gradients, $C_\rho$. A lower value of $C_\rho$
corresponds, in principle, to greater resolution. We compare
\mbox{\emph{RAMSES-8}}, \mbox{\emph{RAMSES-7}}, and \mbox{\emph{RAMSES-9}} (in order
of decreasing $C_\rho$) in the bottom panel of
Fig.~\ref{fig:denspdfram} and right-hand panel of
Fig.~\ref{fig:surfdensgrowthresram}. In this comparison we also
include \mbox{\emph{RAMSES-1}} which effectively has an infinite threshold, i.e. no
grid refinement is permitted based on density gradients. We recall here that
\mbox{\emph{RAMSES-1}} does not make use of grid levels 13 and 14 and
choosing a finite $C_\rho$ allows the adaptive grid to make use of
these levels.

In Fig.~\ref{fig:denspdfram} (lower panel) we note a slight increase
in the maximum density value as $C_\rho$ decreases. The lowest
$C_\rho$ run (\mbox{\emph{RAMSES-9}}) does not follow the trend of increasing peak
density but this is likely linked to it having fewer grid cell than \mbox{\emph{RAMSES-7}}
despite its lower $C_\rho$. We therefore note that the higher density correlates more
with the number of grid cells than with the gradient refinement
threshold and likely does not reflect the ability of this refinement
scheme to place cells in appropriate locations. Examining the inner
annulus in Fig.~\ref{fig:surfdensgrowthresram} (upper right-hand
panel) gives no clear indication of whether the value of $C_\rho$ has any impact
on the arm surface density. The outer annulus (lower right-hand panel
in Fig.~\ref{fig:surfdensgrowthresram}) however suggests that
the lower threshold does enhance the oscillations of the arm
surface density.

We find that, in the context of our galactic disc
simulations, the grid structure is very sensitive to $C_\rho$. We
therefore find that decreasing $C_\rho$ can lead to very little
increase in resolution (because the grid is already refined to an
extent by the Jeans' length criteria), or it can refine the grid to
such a degree that the simulation time increases disproportionately
compared with other approaches. We find that despite the lower value of $C_\rho$ in
\mbox{\emph{RAMSES-9}} than in \mbox{\emph{RAMSES-7}} (which should mean that it
refines grid cells more easily) we actually have fewer grid cells by
the end of the simulation. For the first 150~Myr, the number of grid
cells found in \mbox{\emph{RAMSES-9}} is much higher than in \mbox{\emph{RAMSES-7}},
consistent with its lower refinement threshold, but it then declines
gradually to the value found in Table~\ref{table:parameters}. 

\begin{figure*}
\includegraphics[width=180mm]{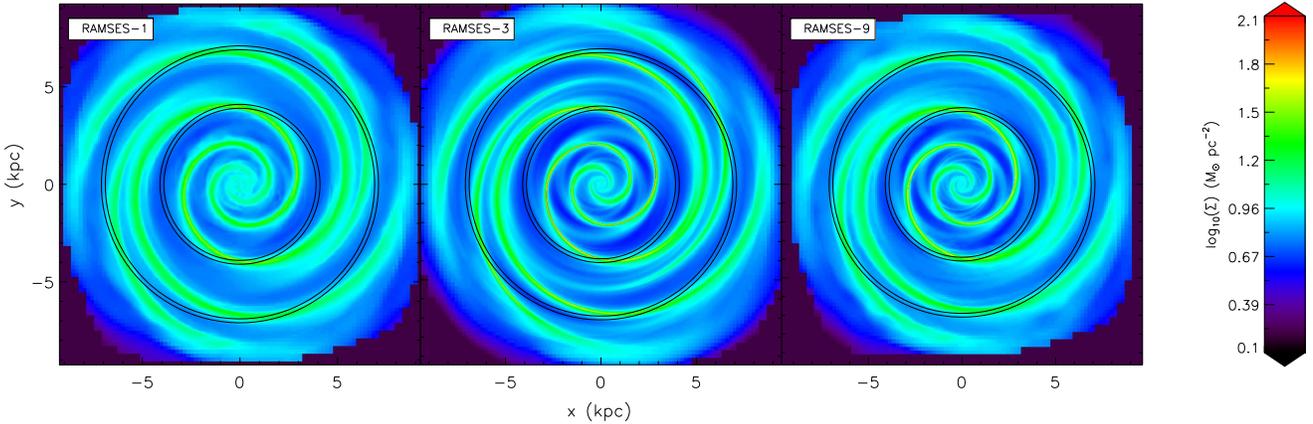}
\caption{Surface density maps for our baseline \textsc{ramses} run (\mbox{\emph{RAMSES-1}}) and for the
  two highest resolution runs using $N_\mathrm{J}$ and $C_\rho$ criteria (\mbox{\emph{RAMSES-3}},
  and \mbox{\emph{RAMSES-9}} respectively) after 250~Myr have passed. Black circles indicate the
  location of the annuli at 4$\pm0.1$~kpc and 7$\pm0.1$~kpc used in our analysis. The most
  significant difference between these runs is the concentration of the arms and the presence
  of more inter-arm structure in the higher resolution runs.}
\label{fig:amrimages}
\end{figure*}

Surface density maps for our baseline \textsc{ramses} run and the two resolution
tests that are most similar to the Lagrangian runs (\mbox{\emph{RAMSES-3}} and \mbox{\emph{RAMSES-9}})
are compared in Fig.~\ref{fig:amrimages}. Both these higher resolution
runs have sharper arms but also exhibit more inter-arm structure. \mbox{\emph{RAMSES-3}},
which resolves the Jean's length with 16 grid cells, presents hints of the rings seen
in the Lagrangian runs and secondary arms that extend further beyond the branching point
(as seen in the Lagrangian runs in Fig.~\ref{fig:discimage}). Similar structures are seen
in \mbox{\emph{RAMSES-9}}, although more feathers are seen near the center and the secondary
arms are not as extended. The emergence of these structures in the higher resolution runs hints that
concordance with the Lagrangian runs is closer.

\section{Discussion and Conclusions}
\label{sec:discconc}

We have examined how an isothermal gas disc evolves under the
influence of an external spiral potential when realised with different
hydrodynamical methods (\textsc{ramses}, \textsc{sphNG}, and \textsc{gizmo}) and as a function of
resolution. With similar resolutions to those found in the literature (and using 
the `acoustic' solver with a MinMod slope limter) we find that our AMR code, \textsc{ramses}, generates a weaker density contrast
between the arm and inter-arm region, less steep vertical profiles and
lower arm surface densities than we see when using the Lagrangian
codes (\textsc{sphNG} and \textsc{gizmo}) as well as less inter-arm structure. When
additional refinement measures are used, \textsc{ramses} generates results in
better agreement with the other codes. If a less diffusive set up is used 
(i.e. an `exact' solver with a MonCen slope limiter) then a measure of similarity 
is also achieved in the resolution of physical structures, vertical density profile, and the density 
PDF. The growth of spiral arm surface density is also enhanced to a similar level as seen in the Lagrangian 
codes but only at smaller radii: arm surface density is still relatively low at greater radii.
In all codes, we also see oscillations in the peak arm surface density, which appear to be
associated with the development of secondary arms, but the
oscillations tend to be very weak with the baseline \textsc{ramses} model. The
\textsc{gizmo} and \textsc{sphNG} codes also display small differences, namely that \textsc{gizmo}
produces the highest densities and surface densities.  

To test why the differences occurred between the codes, we considered
the resolution and viscosity. We found resolution had little effect on
the Lagrangian runs, except for an increasing steepness of the vertical density
profiles with resolution. Whilst artificial viscosity could potentially
effect the results with the \textsc{sphNG} code, our fiducial \textsc{gizmo} runs do
not include artificial viscosity, so this cannot explain the
discrepancies between \textsc{gizmo} and \textsc{ramses}. Furthermore, using \textsc{gizmo} in
`SPH mode' (with constant, \cite{balsara95}, and \cite{cullen10}
viscosity schemes and with pressure-entropy based SPH) we find very
little difference between those runs and the equivalent MFM and MFV
\textsc{gizmo} modes (without viscosity). 
By reducing the artificial viscosity by a factor of 20, the
\textsc{sphNG} models produced more similar results to the baseline
\textsc{ramses} run, however this dramatic change in viscosity greatly
reduces the ability of \textsc{sphNG} to accurately capture shocks.
Thus we believe that the differences between \textsc{ramses}
and the \textsc{sphNG}/\textsc{gizmo} runs, and between \textsc{sphNG} and \textsc{gizmo} are not due to
viscosity. Differences in the choice of smoothing length and the
functional form of the kernel partly account for the differences
between our \textsc{sphNG} and \textsc{gizmo} runs.

We also investigated varying the resolution in the \textsc{ramses} code using a
number of approaches. We first used the most intuitive approach,
increasing the maximum refinement level $\ell_\mathrm{max}$, but surprisingly this
made almost no difference. The reason for this was because the code
was simply not saturating the maximum refinement level. We
stress here though that our simulations are isothermal and do not
include self-gravity, gas cooling, or any sources of forcing other
than our external potential. These processes could drive the density up and
force the grid to refine even without changing the refinement
criteria, though it is not clear that this would improve the
modelling of spiral arms more generally or, for example, the initial
development of Jeans' instabilities.

We secondly tried using refinement criteria based on density
gradients (the kind conventionally employed in idealised code
comparison tests), which we might expect would lead to greater
refinement and better agreement with the arm densities seen with the
other codes. However, we find that the parameters controlling such
refinement schemes are not trivial to select \emph{a priori}. We tend to find
only slight differences in the density PDF and, observe a small
increase in the scale of oscillations in the arm surface density over
time. Our model \mbox{\emph{RAMSES-9}} shows best agreement with the other codes,
but we note that in this run we were refining a significant fraction
of the disc, which leads to an large increase in computing time due to 
refining regions unnecessarily. Again, we note that these
results are true for our particular model choices and hydrodynamical
gradient criteria may be of benefit under other frameworks, e.g. when
supernovae blastwaves are included.

Thirdly we examined increasing refinement by increasing the number of 
grid cells resolving the Jeans' length, $N_\mathrm{J}$. We find
that the greatest, and most reliable, improvement (in the sense of
providing more concordance with other methods) in the density PDF,
vertical density profile and evolution of the surface density is found
by increasing $N_\mathrm{J}$. By changing $N_\mathrm{J}$ from 4 to 16 we find greater
similarities between our AMR and Lagrangian tests. In addition to the
spiral arms, the resolution of the inter-arm regions is also
important, one reason being that dense structures leaving the refined
region should be preserved. To some extent our highest resolution
\textsc{ramses} runs do preserve the inter-arm structures found in the
Lagrangian runs. Although there are still differences between \textsc{ramses}
and the other codes, this finding is in line with other comparison
tests that show more similar results tend to be attained with grid and
particle-based codes \citep{tasker08,kitsionas09,price10} if the resolution is comparable or greater in the
grid code. Our results also seem to concur with the idea in
\cite{price10} that grid-based codes tend to be less effective at
resolving denser regions of simulations.

Our findings that the density in spiral arms differs according to
different numerical codes (we stress that here we mean `code' and not 
`method'), and further that the growth of the maximum
density differs, may have implications for studying star formation in
spiral arms. The different densities may impact the timescales for
gravitational collapse, properties and number of molecular clouds, and
the rate and efficiency of star formation. This may be even more relevant
for simulations with transient spiral arms, whereby the arms come and
go with time, and the time for gas to accumulate into dense structures
in the spiral arms may be fundamentally limited. We have demonstrated
that by refining adaptive grid simulations further than is usual,
similar arm density growth rates are achieved as with Lagrangian
codes. We note that a quasi-Lagrangian refinement scheme could also be
applied in the case where transient spiral arms arise due to a live
stellar disc, but we do not test this idea in this work. We have also not
examined how refinement changes once self-gravity is included, but
likewise leave this to future studies.

To conclude, our key result is that caution is required when justifying the 
use of four grid cells per Jeans’ length as a resolution scheme with AMR 
(or static grid) simulations. This condition is necessary but may not be 
sufficient depending on the hydrodynamical solver.  
For the simulations performed here, the Jeans length needs be 
resolved by at least 16 grid cells to acheive a similar result to that found 
with Lagrangian codes,. The authors are aware of only one study in the 
field of isolated galaxy simulations where the refinement criteria is set 
to a higher value, Petit et al. (2015), wherein 32 cells are used (note the 
authors also applied a quasi-Lagrangian refinement scheme). We find some dependence 
on the solver (in particular using a less diffusive solver also 
produces better resolution), and in general the criteria for the resolution 
may depend on the exact nature of the code.  But given that more diffusive 
set ups are likely frequently used, we believe that this result is an 
important one.  We find that increasing $N_\mathrm{J}$ appears to be the most effective 
and simplest means of increasing the resolution in regions of interest in 
a galactic disc, in particular to study processes such as spiral shocks and 
molecular cloud formation.

\section{Acknowledgments}

CGF, CLD and LK acknowledges funding from the European Research Council for the
FP7 ERC starting grant project LOCALSTAR. We thank the anonymous referee
for a very constructive report. This work used the DiRAC
Complexity system, operated by the University of Leicester IT
Services, which forms part of the STFC DiRAC HPC Facility
(\url{www.dirac.ac.uk}). This equipment is funded by BIS National
E-Infrastructure capital grant ST/K000373/1 and  STFC DiRAC Operations
grant ST/K0003259/1. DiRAC is part of the National
E-Infrastructure. Calculations were also performed on Cray XC30
at the Center for Computational Astrophysics, National Astronomical
Observatory of Japan.

\bibliographystyle{mn2e}
\bibliography{icomp}
\label{lastpage}
\end{document}